\documentclass[showpacs,showemail,preprintnumbers,amsfonts,prb]{revtex4}
\usepackage{graphicx}
\usepackage{bm}
\begin{document}

\title{Low-temperature nucleation in a kinetic Ising model under
different stochastic dynamics with local energy barriers}
\author{Gloria M.\ Buend{\'\i}a$^{1,2,3}$}\email{buendia@usb.ve}
\author{Per Arne Rikvold$^{2}$}\email{rikvold@csit.fsu.edu}
\author{Kyungwha Park$^{4}$}\email{park@dave.nrl.navy.mil}
\author{M.~A.\ Novotny$^{3}$}\email{man40@ra.msstate.edu}
\affiliation{
$^1$Department of Physics, Universidad
Sim{\'o}n Bol{\'\i}var, Caracas 1080, Venezuela\\
$^2$School of Computational Science and Information Technology,
Center for Materials Research and Technology, and Department of Physics,
Florida State University, Tallahassee, Florida 32306-4120, USA\\
$^3$Department of Physics and Astronomy and ERC Center for Computational
Sciences, Mississippi State University, Mississippi State,
Mississippi 39762-5167, USA\\
$^4$Department of Physics, Georgetown University, Washington, DC 20057, USA
}

\date{\today}

\begin{abstract}
Using both analytical and simulational methods, 
we study low-temperature nucleation rates in kinetic Ising lattice-gas models
that evolve under two different Arrhenius dynamics that interpose 
between the Ising states a transition state representing a local energy
barrier. The two dynamics are the transition-state approximation 
[T.~Ala-Nissila, J.~Kjoll, and S.~C.\ Ying, 
Phys.\ Rev.\ B {\bf 46}, 846 (1992)] 
and the one-step dynamic 
[H.~C.\ Kang and W.~H.\ Weinberg, J.\ Chem.\ Phys.\ {\bf 90}, 2824 (1989)]. 
Even though they both obey detailed balance and are here applied to a 
situation that does not conserve the order parameter, 
we find significant differences between the nucleation rates observed
with the two dynamics, and between them and the standard Glauber dynamic 
[R.~J.\ Glauber, J.\ Math.\ Phys.\ {\bf 4}, 294 (1963)], 
which does not contain transition states. 
Our results show that great care must be
exercised when devising kinetic Monte Carlo transition rates for
specific physical or chemical systems. 
\end{abstract}


\maketitle

\section{Introduction}
\label{sec:I}
Nucleation phenomena are crucially important in many scientific disciplines, 
ranging from biochemistry\cite{ONUC} to earth sciences,\cite{LASA98} 
astrophysics,\cite{PATZ} and cosmology,\cite{KAST98,KAST00}
and they have been studied by kinetic Monte Carlo simulations in
electrochemistry,\cite{BROW99A,MITC02,BERT04A,BERT04B} 
materials science,\cite{COMB00,AUER01,FICH02}
magnetism,\cite{RICH94,NOVO02A} and atmospheric science,\cite{MAHN03} 
to mention just a few.
However, many questions in nucleation theory are not yet resolved, and recently
there has been much interest in using kinetic Ising or lattice-gas systems as
theoretical and computational models of nucleation.  
In particular, over the last decade much work has been done on their
dynamical behavior at very low temperatures, both in the 
computational\cite{NOVO97,NOVO03,NOVO02,PARK02B,PARK02C,PARK02,SHNE02,%
SHNE03,SHNE03B} and 
mathematical\cite{NEVE91,SCHO92,SCOP93,SCOP93A,SCHO94,KOTE94,OLIV95,%
DEHG97,DEHG97B,BOVI02}
physics communities. 
While this body of literature is divided between the Ising spin and lattice-gas 
formulations, we here choose to use the more symmetric Ising language. 
However, we remind the reader that the Ising spin up and down states correspond 
to lattice-gas filled and empty sites, and that the magnetic field in the 
Ising model is proportional to a chemical potential in the lattice-gas case. 
The explicit mappings are straightforward and can be found in 
Endnote~\onlinecite{NOTE1}. 

In a typical numerical nucleation experiment, the system is prepared in a
metastable state with all spins antiparallel to the applied field, and is
then allowed to evolve under a stochastic dynamic until the 
magnetization reaches zero on its way to the stable phase consisting of spins
mostly aligned with the field. 
In the magnetic (ferroelectric) interpretation, this
situation corresponds to magnetization (polarization) switching in an applied 
field,\cite{RICH94,NOVO02A} while in the lattice-gas interpretation it would
correspond to submonolayer adsorption in the limit that lateral
adsorbate diffusion can be ignored.\cite{RAMO99,NOVO00} 

At very low temperatures, the dynamical
behavior is influenced by the discreteness of the lattice, and it is
possible to calculate exactly both the critical droplet (the saddle-point
configuration) and the most probable path the system follows through 
configuration
space during a nucleation event. The average nucleation time, defined as the 
inverse of the nucleation rate per unit system volume, has,
for a two-dimensional system evolving under the Metropolis dynamic
\cite{METR53,LAND00}
in the low-temperature limit, been rigorously shown to be\cite{NEVE91} 
\begin{equation}
\langle \tau \rangle = A e^{\beta \Gamma} .
\label{eq:tau}
\end{equation}
Here, $A$ is a nonexponential prefactor, 
$\Gamma$ is related to the energy barrier 
separating the metastable state from the stable state,
and $\beta = 1/ k_{\rm B} T$ where $T$ is the temperature and $k_{\rm
B}$ is Boltzmann's constant (hereafter set equal to one). With this
definition, $\langle \tau \rangle$ is independent of the system size.  

In kinetic Monte Carlo simulations, time is usually measured in 
Monte Carlo steps per site
(MCSS), corresponding to one attempted update per site in the system. Since
thermal fluctuations occur in parallel all over the system, this is the
measure of Monte Carlo time that would be proportional to physical time. At 
the very low temperatures and relatively small system sizes considered here, the
average waiting time for a nucleation event is much longer than the time it
would take a supercritical droplet to grow to a size comparable to the size
of the system. As a result, the change of the order parameter from its
metastable to its stable value is accomplished by {\it one single droplet\/}
-- the first one to nucleate after the simulation is started.  
Since nucleation events can occur independently all over the system, the
metastable lifetime in this single-droplet regime, operationally defined as
the time until the system magnetization reaches zero and measured in MCSS, 
is inversely proportional to the system 
volume.\cite{RIKV94A,RIKV94,PARK02,PARK02B}
However, the quantity of interest here is the inverse nucleation rate
$\langle \tau \rangle$, which is independent of the system size. To remove
the volume dependence from the time measurements and obtain data directly
applicable to $\langle \tau \rangle$, we therefore here choose to measure
simulation times in units of individual Monte Carlo steps (MCS). 
It is in these time
units that we shall loosely refer to $\langle \tau \rangle$ as ``the
metastable lifetime.'' Lifetimes measured in the
conventional MCSS units can be obtained through division by the number of
sites in the system, $N$. 

Neves and Schonmann\cite{NEVE91} have shown that the critical droplet is  
a rectangle of overturned spins of size $\ell \times ( \ell - 1)$
with an extra ``knob'' consisting of one overturned spin on one of its 
long sides. The critical length $\ell$ is given by 
\begin{equation}
\ell = \lfloor 2 J / |H| \rfloor + 1 
\;,
\label{eq:ell}
\end{equation}
where $\lfloor x \rfloor$ denotes the integer part of $x$. Here, $J>0$ is the 
nearest-neighbor interaction constant of the Ising model, which will
henceforth be set equal to unity, and $H$ is the applied field in the 
Ising model. The critical length thus
changes discontinuously at values of $|H|$ such that $2 / |H|$ is an integer. 

It has been commonly assumed that the quantity $\Gamma$ in
Eq.~(\ref{eq:tau}) equals the height of the energy barrier separating the
metastable state from the equilibrium state. However, 
we recently showed that $\Gamma$ actually depends on the specific stochastic
dynamic under which the system evolves.\cite{PARK04} 
Only for certain dynamics, which
include the Metropolis dynamic used in Ref.~\onlinecite{NEVE91}, as well as
the commonly used standard Glauber dynamic,\cite{GLAU63,LAND00} does
$\Gamma$ actually equal the energy barrier.\cite{NOVO03,PARK04} 
For other dynamics that also satisfy detailed balance,
and so are perfectly admissible from the point of view of eventually
bringing the system to equilibrium, $\Gamma$ differs from the energy 
barrier.\cite{PARK04} 

However, neither of the dynamics studied in Ref.~\onlinecite{PARK04} contains a
microscopic energy barrier against individual spin flips. Since such a
barrier is often needed in modeling the dynamics of physical systems,
we here consider two specific dynamics that both possess a microscopic barrier.
These are the transition dynamics approximation (TDA), introduced by
Ala-Nissila, et al.,\cite{ALAN92A,ALAN92,ALAN02} 
and the commonly used one-step dynamic (OSD).\cite{COMB00,FICH02,KANG89,FICH91}
Both can be used in studies of adsorption/desorption and diffusion processes. 
Formal definitions of these transition rates are given in Sec.~\ref{sec:M}. 

The rest of this paper is organized as follows. 
The Ising lattice-gas model and the different stochastic Monte Carlo
dynamics are introduced in Sec.~\ref{sec:M}, and the methods used and
results obtained are discussed in Sec.~\ref{sec:R}. In particular, 
analytical calculations using a one-step Markov chain are discussed in
Sec.~\ref{sec:onestep}, computer-aided analytical calculations using
absorbing Markov chains are discussed in Sec.~\ref{sec:fa}, and Monte
Carlo simulations using the Monte Carlo with absorbing Markov chains
method are discussed in Sec.~\ref{sec:MC}. A summary and
conclusions are given in Sec.~\ref{sec:conc}. 
Some preliminary analytical results on the TDA dynamic are contained in
Ref.~\onlinecite{BUEN04}. 

\section{Model and Dynamics}
\label{sec:M}
The square-lattice $S=1/2$ Ising ferromagnet with unit interaction 
is defined by the Hamiltonian 
\begin{equation}
{\cal H} = - \sum_{\langle \alpha,\beta \rangle} \sigma_\alpha \sigma_\beta 
- H \sum_{\alpha} \sigma_\alpha \;,
\label{eq:ham}
\end{equation}
where the Ising spins $\sigma_\alpha = \pm 1$, 
$H$ is the dimensionless 
applied magnetic field, $\sum_{\langle \alpha,\beta \rangle}$ runs
over all nearest-neighbor bonds on a square lattice, and $\sum_{\alpha}$ runs
over all lattice sites. (The mapping to the lattice-gas formulation
is given in Endnote~\onlinecite{NOTE1}.)
When this system develops under the standard
Glauber dynamic with spin-flip rate\cite{GLAU63,LAND00}
\begin{equation}
W_{\mathrm G} = \frac{1}{1+\exp{(\beta \Delta E)}}
\;,
\label{eq:G}
\end{equation}
where $\Delta E$ is the energy change that would result from the
transition, $\Gamma$ in Eq.~(\ref{eq:tau}) is given
by\cite{NEVE91,NOVO03}
\begin{equation}
\Gamma_{\rm G} = 8 \ell - 2 |H| ( \ell^2 - \ell + 1 ) \;.
\label{eq:gamma}
\end{equation}
There is agreement in the literature that for $1 < |H| < 2$, the 
prefactor is given by $A_{\rm G} = 3/[8 ( \ell - 1)]$.\cite{BOVI02} 
{}For $|H| < 1$ and such that $2/|H|$ is not an integer,
Ref.~\onlinecite{BOVI02} also gives this result, while 
Refs.~\onlinecite{SHNE03B,SHNE03} instead give 
$A_{\rm G} = 3/[8 ( \ell - 1) + 4]$. 
The extra factor of 4 in the denominator in the latter result would be 
due to the effect of degenerate configurations with the same energy and 
the same number of overturned spins that occur 
for $\ell \ge 3$.\cite{SHNE03} 
Since we in this paper perform
numerical calculations only for $|H| > 1$ ($\ell \le 2$), 
we have not explicitly tested this discrepancy. 
Results for $|H| \ge 2$ were obtained in
Refs.~\onlinecite{PARK04,SHNE02,SHNE03B,SHNE03}. 
Note that $\Gamma_{\rm G}$ is a continuous function of
$|H|$, even though $\ell$ and $A_{\rm G}$ are
discontinuous wherever $2/|H|$ is an integer. 

The interpretation of $\Gamma$ as given in Eq.~(\ref{eq:gamma}) is simply the
energy difference between the metastable state and the critical droplet. 
In Ref.~\onlinecite{PARK04} we showed that there exist stochastic
dynamics that satisfy detailed balance, as does the Glauber 
dynamic, but for which $\Gamma$ is different from the value given by
Eq.~(\ref{eq:gamma}), even though the critical droplet 
remains unchanged by the change of dynamic. The specific dynamic for which
this was demonstrated was the ``soft Glauber dynamic,''\cite{RIKV02} for
which the transition rates are given by 
\begin{equation}
W_{\rm SG} = 
\frac{1}{1+\exp{(\beta \Delta E_J)}} \,
\frac{1}{1+\exp{(\beta \Delta E_H)}}
\;,
\label{eq:SG}
\end{equation}
where $\Delta E_J$ and $\Delta E_H$ are the changes in the 
interaction energy and in the field energy, respectively.
Consequently, the total energy change is $\Delta E = \Delta E_J + \Delta E_H$. 
It is the factorization property into one part that depends only on 
$\Delta E_J$ and another that depends only on $\Delta E_H$ that defines this
dynamic as ``soft."\cite{MARR99,RIKV00B,RIKV02} In contrast, the standard
Glauber dynamic, Eq.~(\ref{eq:G}), which does not have this factorization
property, is described as ``hard.'' 
The need to use soft dynamics in cases where the `field' represents a
chemical-potential difference has been recognized in some Monte Carlo 
simulations of crystal growth.\cite{GUO90,KOTR91,HONT97}

\begin{figure}[ht]
\includegraphics[angle=0,width=.47\textwidth]{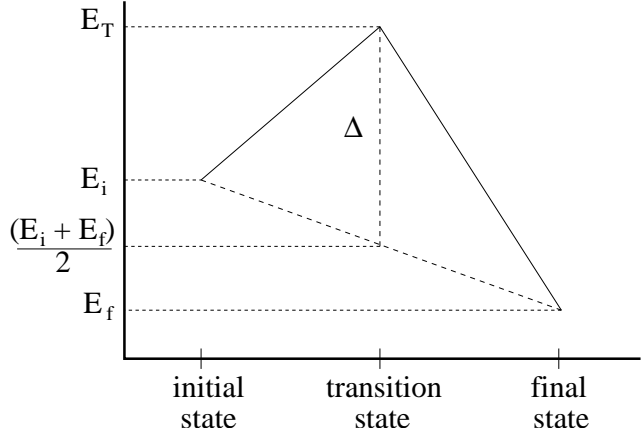}
\vspace{.3cm}
\caption[]{
Schematic picture of the transition barrier in the symmetric Butler-Volmer
approximation, used to calculate the TDA and OSD transition rates. 
}
\label{fig:BV}
\end{figure}

In many applications it is physically reasonable to introduce a
microscopic transition barrier.\cite{ALAN92A,MITC02,BERT04A,BERT04B}
Such a barrier is not part of the Ising
or lattice-gas Hamiltonian, and it is therefore absent in  
the transition rates discussed above. Rather, the barrier represents a
transition state which is inserted between the states allowed in the
Hamiltonian, such as a saddle point in a corrugation potential for
particle diffusion, or a high energy associated with a transitional 
spin state that is not along one of the two directions allowed by the
Ising Hamiltonian. 
Dynamics that contain such a barrier are often called Arrhenius dynamics, 
and they can be viewed as a way to use the lattice-gas or Ising model to
simulate dynamics in an underlying, continuous potential.\cite{ALAN92A,MITC02}
Here we use the following approximation  
to express the transition-state
energy:\cite{KANG89,FICH91,ALAN92A,ALAN92}
\begin{equation}
E_{\rm T} = \frac{E_{\rm i} + E_{\rm f}}{2} + \Delta
\;,
\label{eq:BV}
\end{equation}
where $E_{\rm i}$ and $E_{\rm f}$ are the
initial and final energies, and $\Delta$ is the bare, microscopic 
energy barrier. 
In electrochemical applications, such as electron- or ion-transfer
reactions, this corresponds to the 
symmetric Butler-Volmer approximation.\cite{BOCK70,SCHM96,BROW99A} 
The construction corresponding to 
Eq.~(\ref{eq:BV}) is illustrated in Fig.~\ref{fig:BV}. 

The two specific dynamics considered in this paper are the TDA and OSD 
rates with the transition-state energy $E_{\rm T}$ given by Eq.~(\ref{eq:BV}). 
They are defined as\cite{ALAN92A,ALAN92,ALAN02}  
\begin{equation}
W_{\rm TDA} = 
\frac{1}{1+\exp[\beta ( E_{\rm T} - E_{\rm i})]} \,
\frac{1}{1+\exp[\beta ( E_{\rm f} - E_{\rm T} )]}
\;,
\label{eq:TDA}
\end{equation}
and\cite{KANG89,FICH91} 
\begin{equation}
W_{\rm OSD} 
=
\exp [- \beta( E_{\rm T} - E_{\rm i})]
=
\exp (-\beta \Delta ) \,
\exp \left(- \beta \frac{E_{\rm f} - E_{\rm i}}{2} \right)
\;,
\label{eq:OSD}
\end{equation}
respectively. From the definitions of hard and soft dynamics given above, we
see that the TDA dynamic cannot be factored into parts that depend only
on $\Delta E_J$ and $\Delta E_H$ and thus is hard. 
On the other hand, the OSD dynamic can be factored in this way and 
thus is soft. 
While these dynamics are often used for diffusion problems
(i.e., conserved order parameter),\cite{ALAN92A,ALAN92,ALAN02} 
we will here use them in the simpler context of magnetization switching
or adsorption/desorption (i.e., nonconserved order parameter). 

\section{Results}
\label{sec:R}
Like we did in Ref.~\onlinecite{PARK04}, in this paper 
we obtain our results in three different ways. First, we calculate 
analytically by hand the first-passage time from the metastable state
to an absorbing state just beyond the saddle point 
in an approximation that the path in configuration
space corresponds to a simple one-step Markov process.\cite{KAMP92} Second,
we perform computer-aided analytic calculations  
using the technique of absorbing Markov chains 
(AMC),\cite{NOVO97,IOSI80,NOVO01} 
allowing for multiple branching paths and ``blind alleys'' 
on the way to the absorbing state(s).
Third, we perform Monte Carlo simulations using the Monte Carlo with AMC
(MCAMC) technique.\cite{NOVO02,NOVO95,NOVO01,NOVO03B}
The first method provides the clearest physical insight, and for noninteger 
values of $2/|H|$ the results are fully confirmed by the other two. 

\subsection{One-step Markov chain}
\label{sec:onestep}
\begin{figure}[ht]
\includegraphics[angle=0,width=.47\textwidth]{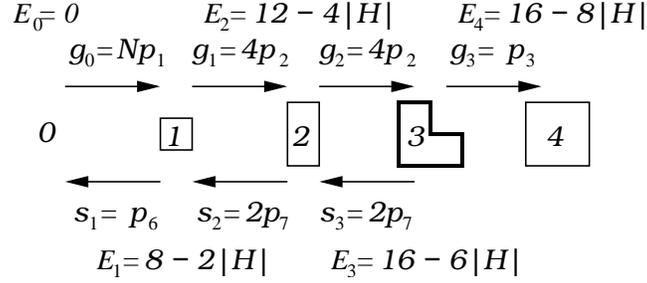}
\vspace{.3cm}
\caption[]{
The states in the one-step 
Markov chain of clusters of $i = 0,...,4$ overturned spins, 
used to calculate the metastable lifetime $\langle \tau \rangle$ 
analytically by hand. The right-pointing 
arrows give the growth rates $g_{i-1}$, and the left-pointing arrows give the 
shrinkage rates $s_i$ for $i =1$, 2, and~3. The numbers preceding the
spin-flip rates $p_m$ (see Table~\protect\ref{tab:prob}) 
are the degeneracy factors $n_{i-1}^g$ and $n_i^s$ for $g_{i-1}$ 
and $s_i$, respectively. The energies $E_i$ (relative to 
the metastable state, $i=0$) are given at the top of the figure for even $i$ 
and at the bottom for odd $i$. 
See discussion in the text. 
From Ref.~\protect\onlinecite{PARK04}. 
}
\label{fig:conf}
\end{figure}

As we discussed in Ref.~\onlinecite{PARK04}, 
the one-step Markov chain for $1 < H < 2$ ($\ell = 2$) 
corresponds to the configurations labeled $i = 0,...,4$ in
Fig.~\ref{fig:conf}. The label $i$ corresponds to the number of overturned 
spins, such that the starting configuration with all spins up has
$i=0$, and the saddle point has $i = i^* =3$. In general the
absorbing state is labeled $I \ge i^* \ge 1$. The mean 
time spent in state $i$ (the sojourn time) is denoted  
$h_i$, the rate at which the cluster of overturned spins grows
from $i$ to $i+1$ is denoted $g_i$, 
and the rate with which it shrinks from $i$ to $i-1$ is denoted $s_i$. These
quantities satisfy the equation\cite{KAMP92,KOLE98,KOLE98B,NOVO99B}
\begin{equation}
h_{i-1} = (s_i h_i + N )/ g_{i-1}
\label{eq:Miro}
\end{equation}
with boundary conditions $s_I = s_0 = 0$. 
The number $N$ of sites in the system represents the total
probability current through the Markov chain.\cite{KAMP92} 
This relation enables us to obtain
$h_i$ recursively as $h_{I-1} = N/g_{I-1}$ and 
\begin{equation}
h_i = \frac{N}{g_i} + \sum_{k=1}^{I-1-i} \frac{N}{g_{i+k}} 
                      \prod_{j=1}^k \left( \frac{s_{i+j}}{g_{i+j-1}} \right)
\label{eq:Miro2}
\end{equation}
for $0 \le i \le I-2$. Assuming that $I > i^*$ and that the growth 
time for the supercritical droplet can be neglected in comparison with the 
nucleation time, the metastable lifetime $\langle \tau \rangle$ 
is given by the mean first-passage time to $I$, 
$\langle \tau \rangle = \langle \tau_I \rangle = \sum_{i=0}^{I-1} h_i$. 
Grouping the terms in the sum according to the ``unpaired'' factors
$1/g_{i+k}$ in Eq.~(\ref{eq:Miro2}), we get
\begin{equation}
\langle \tau_I \rangle 
=
\frac{N}{g_0} + \sum_{l=1}^{I-1} \frac{N}{g_l} 
\left( 1 + \sum_{k=1}^l 
\prod_{j=0}^{k-1} \frac{s_{l-j}}{g_{l-j-1}} 
\right) 
\;.
\label{eq:Miro3}
\end{equation}
With a suitable interpretation of the factor $N$, 
this result is general for any one-step Markov chain with absorption at
$I$, regardless of the values of $g_i$ and $s_i$. However, for the Ising
or lattice-gas 
model the transition probabilities are related by detailed balance, so that 
\begin{equation}
\frac{s_i / n^s_i}{ g_{i-1} / n^g_{i-1}} = 
e^{\beta (E_i - E_{i-1})} 
,
\label{eq:detbal}
\end{equation}
where $E_i$ is the energy of state $i$. 
The degeneracy factors $n^s_i$ and $n^g_{i-1}$ are the
numbers of lattice sites at which a single spin flip can shrink the
cluster from $i$ to $i-1$ and analogously for growth
from $i-1$ to $i$, respectively. 
As a result, Eq.~(\ref{eq:Miro3}) becomes 
\begin{equation}
\langle \tau_I \rangle 
=
\frac{N}{g_0} + \sum_{l=1}^{I-1} \frac{N}{g_l} 
\left( 1 + \sum_{k=1}^l 
e^{\beta (E_l - E_{l-k})} 
\prod_{j=0}^{k-1} \frac{n^s_{l-j}}{n^g_{l-j-1}} 
\right) 
\;.
\label{eq:Miro4}
\end{equation}
The number of sites that can give rise to a transition from zero
to one overturned spin equals the system size $N$, so $n_0^g = N$. In
contrast, the other $n_i^g$ and $n_i^s$ are just functions of the cluster's
size ($i$) and its shape, and thus independent of $N$. Consequently, 
each term in Eq.~(\ref{eq:Miro4}) that has a factor of $g_0$ or
$n_0^g$ in the denominator is {\it independent\/} of $N$. 
This turns out to be the first term, $N/g_0$, as well as exactly one term
for each value of $l$ in the outer sum, namely, the one corresponding
to $k=l$ in the inner sum. All the other terms are of $O(N)$.
We note that the $N$-independent terms in Eq.~(\ref{eq:Miro4})
constitute a series analogous to the expression for the
inverse nucleation rate, given in Eq.~(15) of Ref.~\onlinecite{SHNE03}.
In the limit of $\beta \rightarrow \infty$, Eq.~(\ref{eq:Miro4}) is dominated
by the term or terms with the largest exponential factor. 
The selection of these terms, which determine both $\Gamma$ and $A$,
is described below. However, we first need to
determine the low-temperature spin-flip rates in the two different dynamics 
considered here. 

For the square-lattice Ising system, all possible
spin configurations can be classified 
into 10 classes, determined by the value $\sigma$ of a spin 
($+$ for the metastable direction and $-$ for the stable direction) 
and the number $N_+$ of its nearest neighbors that are in the metastable 
direction.\cite{BORT75} 
In the low-temperature limit, the rate $p_m$ for flipping 
a spin in class $m$ ($m=1,2,...,10$) [given by Eqs.~(\ref{eq:TDA})
and~(\ref{eq:OSD}) for the TDA and OSD dynamics, respectively], 
can be simplified as shown in Table~\ref{tab:prob}. 

Figure~\ref{fig:conf} corresponds to a one-step
Markov chain with $I=4$. For $1 < |H| < 2$ the 
saddle-point configuration ($i=i^*$) 
is the L-shaped three-site cluster with $\ell = 2$ ($i^*=3$).
Among the dominant terms in Eq.~(\ref{eq:Miro4})
is always the term with $k = l = i^* -1$. 
For all $H < 2$, the growth from $i^*-1$ 
to $i^*$ always involves adding a ``knob'' to the long side of an $\ell
\times (\ell -1)$ rectangle, such that 
$g_{i^* -1} = n_{i^* -1}^g p_2 = 2 \ell p_2$.
For $2<|H|<4$, the saddle point is a single overturned spin 
($\ell = 1$), so $i^* =1$
and $g_{i^* -1} = g_0 = n_0^g p_1 = N p_1$. 
To avoid $N$-dependent growth terms in Eq.~(\ref{eq:Miro4}), 
for $|H|>4$ the lifetime must be defined as the first-passage time to one
overturned spin, so that $\langle \tau \rangle = \langle \tau_1 \rangle 
= N/g_0 = 1/p_1$.

To obtain $\Gamma$ and $A$ in the two
dynamics for $|H| > 1$, we explicitly write out Eq.~(\ref{eq:Miro4}) 
for $I=4$ with the $s_i$, 
$g_{i-1}$, and $E_i$ obtained from Fig.~\ref{fig:conf}:
\begin{eqnarray}
\langle \tau_4 \rangle 
&=&
{1 \over p_1}  
+ {1 \over 4 p_2 } \left( N + e^{\beta(8 - 2|H|)} \right)      
+ {1 \over 4 p_2 } \left( N + {N \over 2} e^{\beta(4 - 2|H|)} 
                            + {1 \over 2} e^{\beta(12 - 4|H|)} \right)
                                                                   \nonumber\\
&&+ {1 \over p_3 } \left( N + {N \over 2} e^{\beta(4 - 2|H|)} 
                            + {N \over 4} e^{\beta(8 - 4|H|)} 
                            + {1 \over 4} e^{\beta(16 - 6|H|)} 
			    \right)                                \nonumber\\
&\equiv& {\cal A} + {\cal B} + {\cal C} + {\cal D} 
\;.
\label{eq:tauI4}
\end{eqnarray}
Here, ${\cal A}$ is the first-passage time from $i=0$ to~1, 
${\cal B}$ is the first-passage time from $i=1$ to~2, etc.
Using the values of
$p_m$ from Table~\ref{tab:prob} we can identify the dominant 
terms in $\langle \tau_4 \rangle$ and explicitly obtain $A$
and $\Gamma$ for both dynamics for all $|H|>1$. 
To avoid complicating overlaps of different field regimes, in the remainder of
this paper we restrict ourselves to $0 \le \Delta < 1$.

\begin{figure}[ht]
\includegraphics[angle=0,width=.47\textwidth]{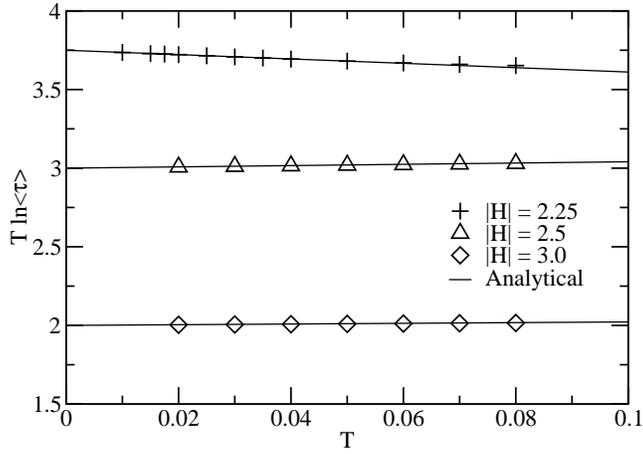}
\vspace{.3cm}
\caption[]{
Analytical (lines) and simulated (data points) results for 
$T \ln \langle \tau \rangle \approx \Gamma + T \ln A$ for the 
TDA dynamic, shown vs $T$ for three different values of $|H|$. 
Error bars, of the order of $T/\sqrt{2000}$, are too small to be seen on
the scale of this figure. 
Weighted two-parameter linear fits to the simulation data yield the simulated
estimates for $A$ and $\Gamma$, given in Table~\protect\ref{tab:TDAAG} 
and shown in Fig.~\protect\ref{fig:TDA}. 
}
\label{fig:all}
\end{figure}

\begin{figure}[ht]
\includegraphics[angle=0,width=.47\textwidth]{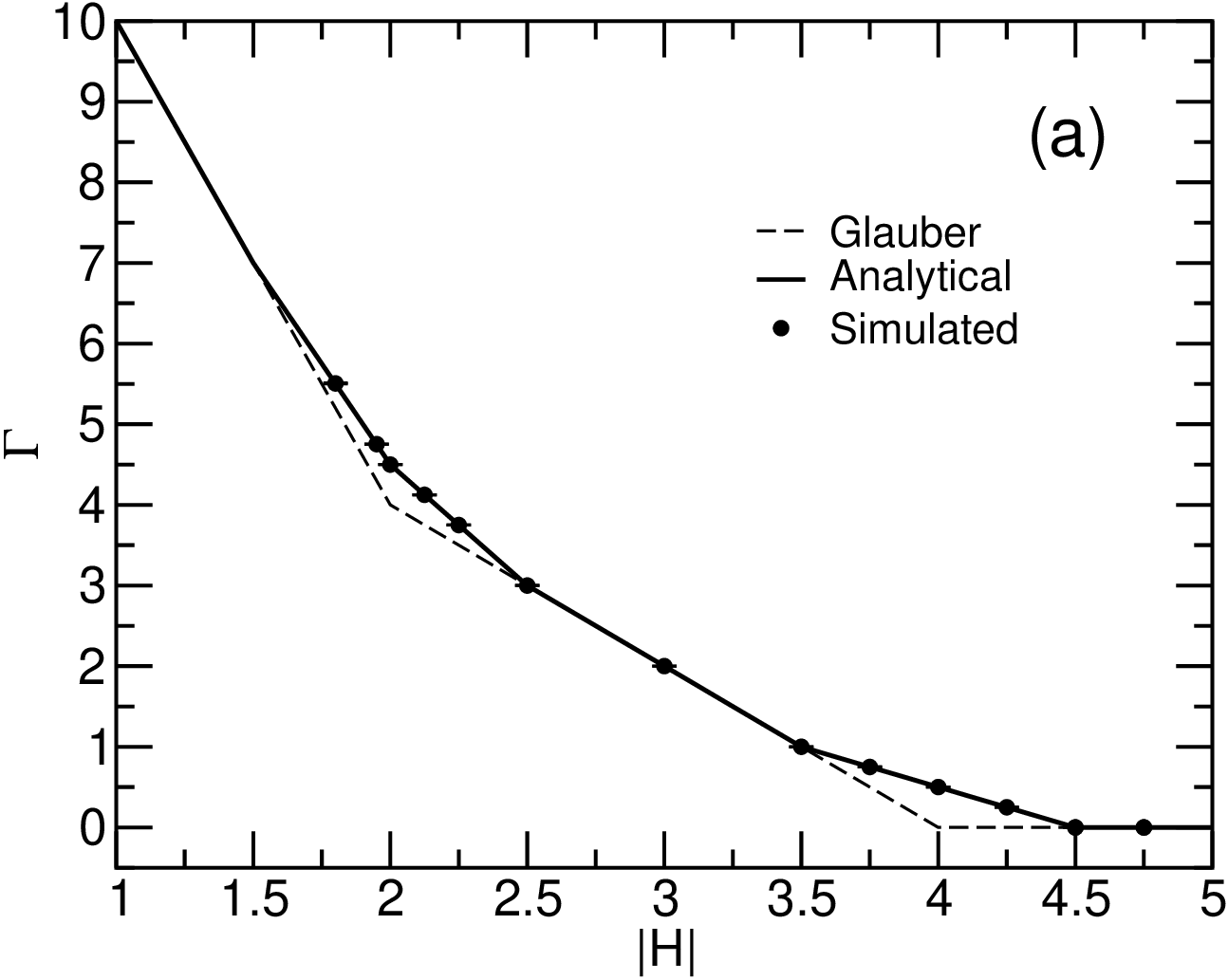}
\hspace{0.5truecm}
\includegraphics[angle=0,width=.47\textwidth]{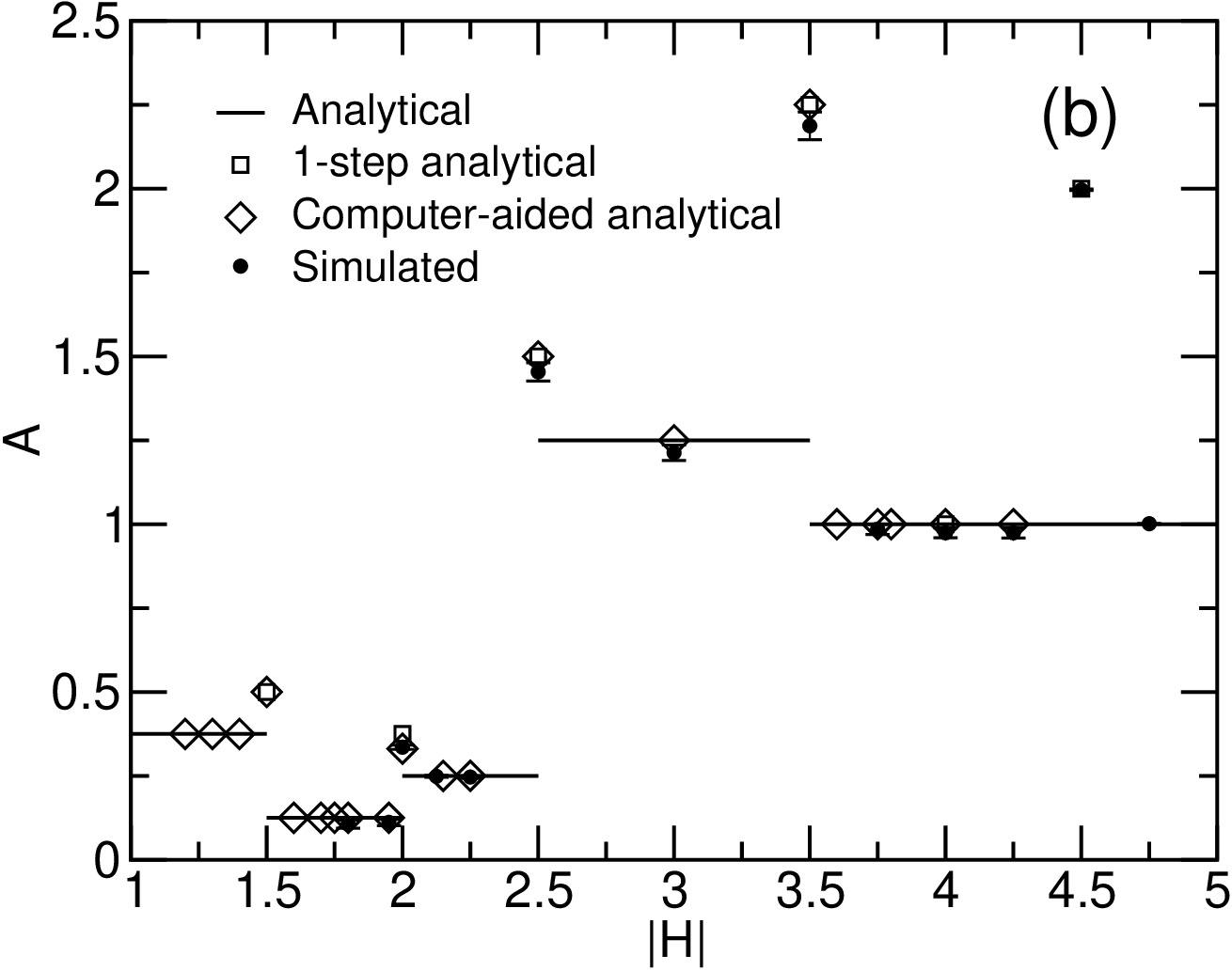}
\vspace{.3cm}
\caption[]{
Analytical and simulated results for $\Gamma$ (a) and $A$ (b) vs $|H|$ for
the TDA dynamic, using $\Delta = 0.5$. Results for $\Gamma$ for
the standard Glauber dynamic (from Ref.~\protect\onlinecite{PARK04}) 
are shown by dashed lines in (a). 
See discussion in the text. 
}
\label{fig:TDA}
\end{figure}

{\it TDA dynamic\/}: 
For all $|H|<4$, $\langle \tau_4 \rangle$ 
is dominated by one or more of the $N$-independent terms. 
For $1 < |H| \le 2 - \Delta$ the dominant part is $\cal D + C$, 
for $2-\Delta < |H| < 2$ it is $\cal C$, 
for $|H|=2$ it is $\cal C + B$,
for $2 < |H| < 2 + \Delta$ it is $\cal B$,
for $2 + \Delta \le |H| \le 4 - \Delta$ it is $\cal B+A$, and 
for $4 - \Delta < |H| < 4$ it is $\cal A$. 
As mentioned above, 
for $|H| \ge 4$ the system is unstable, so the terms $\cal B ... D$,
whose dominant parts are then of $O(N)$, correspond to $\it growth$ of 
a supercritical droplet, 
rather than nucleation. Consequently, the metastable lifetime must be
redefined as the first-passage time to a {\it single\/} overturned spin, 
$\langle \tau \rangle = \langle \tau_1 \rangle = {\cal A}$, which is
still independent of $N$. 
The corresponding results for $\Gamma_{\rm TDA}$ and $A_{\rm TDA}$ 
for all values of $|H| > 1$ are listed in Table~\ref{tab:TDAAG} and 
shown in Figs.~\ref{fig:all} and~\ref{fig:TDA}. 

{\it OSD dynamic\/}: 
For all $|H|<4+\Delta$, $\langle \tau_4 \rangle$ 
is dominated by one or more of the $N$-independent terms. 
For $1 < |H| < 2$ the dominant part is $\cal C$, 
for $|H|=2$ it is $\cal C + B$,
for $2 < |H| < 3$ it is $\cal B$,
for $|H|=3$ it is $\cal B + A$,
and for $3 < |H| < 4 + \Delta$ it is $\cal A$. 
In the same way as for the TDA dynamic for $|H|>4$, 
for $|H| \ge 4 + \Delta$, the 
$O(N)$ growth terms in $\cal B ... D$ would dominate the nucleation
term $\cal A$ and must be removed by defining $\langle \tau \rangle 
= \langle \tau_1 \rangle = {\cal A}$. 
The corresponding results for $\Gamma_{\rm OSD}$ and $A_{\rm OSD}$ 
for all values of $|H| > 1$ are listed in Table~\ref{tab:BVAG} and 
shown in Fig.~\ref{fig:OSDgam}.

\subsection{Computer-aided analytic calculation}
\label{sec:fa}
\begin{figure}[ht]
\includegraphics[angle=0,width=.47\textwidth]{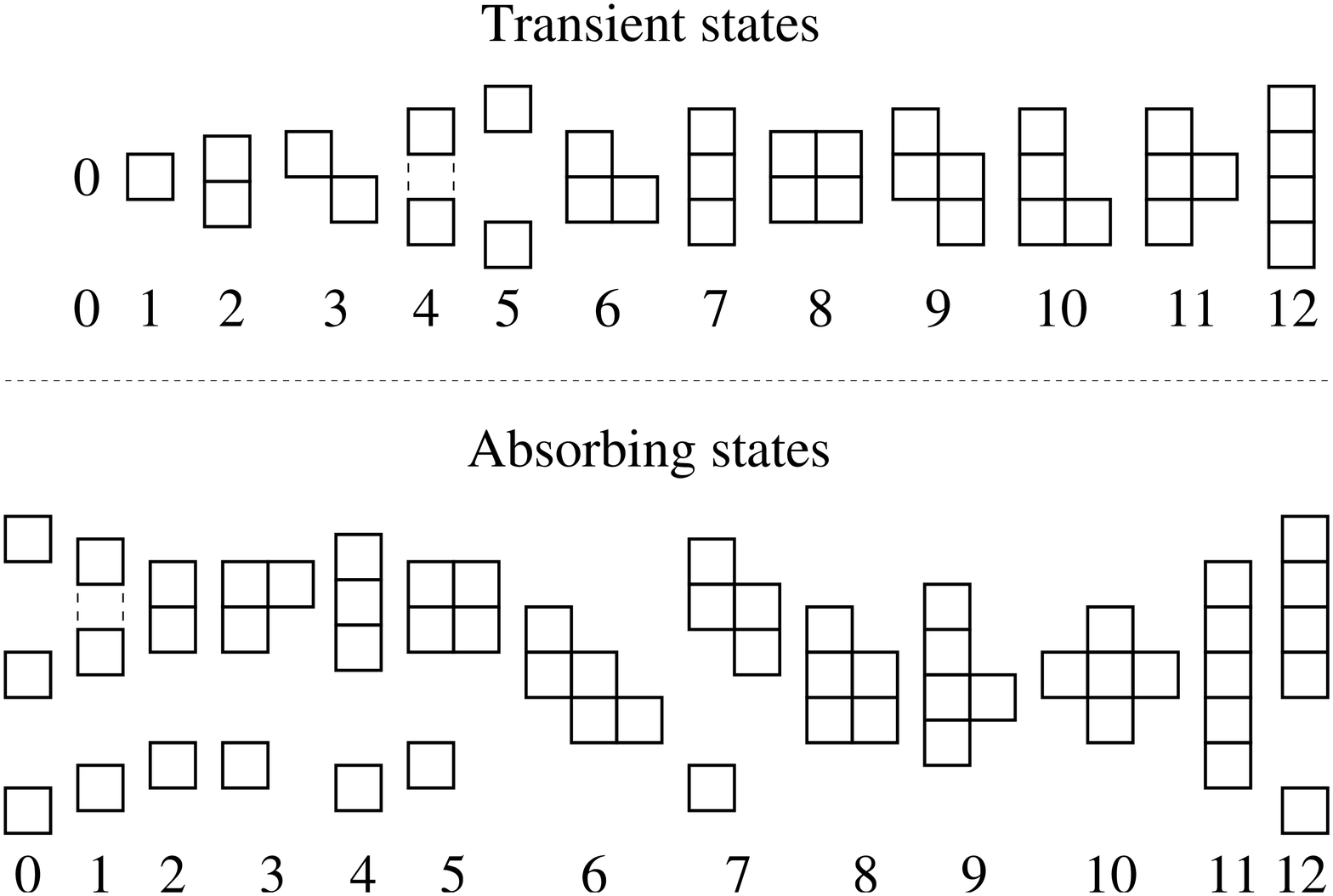}
\vspace{.3cm}
\caption[]{
The transient (T) and absorbing (A) states used in the full AMC calculations
with 13~T and 13~A states. Squares denote sites with
an overturned spin. For states that do not consist of a contiguous
cluster of nearest-neighbor sites, connecting dashed
lines indicate that the overturned spins are separated by a single site 
[T state~4 (T4) and A state~1 (A1)]. 
In the noncontiguous state T5, the two overturned spins 
can be at any sites that are further apart than third-nearest neighbors.
Among the A states, noncontiguous sites that are not connected
by dashed lines can be in any relative positions, except
nearest-neighbor. Several of the A states can also have other
shapes than the ones shown here. For example, the four-site cluster in
A7 can also be L-shaped like T10, and the low-symmetry 
five-site clusters A6 and A9 can have several different shapes. 
In addition, all states also include configurations obtained from
those shown by rotation or reflection. 
All the A states, including variations and states equivalent under
symmetry, were taken into account in the calculation. 
}
\label{fig:full}
\end{figure}

To check the results of the one-step Markov-chain approximation, 
analytic calculations of $\langle \tau \rangle$ were also performed by the 
AMC method,\cite{IOSI80,NOVO01} using $t=13$ transient states and 
all the $a$ absorbing states that can be reached from the transient subspace
by a single spin flip. The transition matrix defining the AMC has
the form
\begin{equation}
\label{EAMC0}
{\bf M}_{(a+t)\times(a+t)} =
\pmatrix{
{\bf I}_{a\times a} & {\bf 0}_{a\times t} \cr
{\bf R}_{t\times a} & {\bf T}_{t\times t} \cr
}
\;,
\end{equation}
where the sizes of the blocks have been explicitly indicated. Here,
${\bf I}$ is an identity matrix, ${\bf 0}$ is a matrix with
all zero elements, ${\bf T}$ is called the transient matrix,
and ${\bf R}$ is called the recurrent matrix. The latter two matrices contain
the transition rates within the transient subspace and from the transient to
the absorbing subspace, respectively. 
We use the convention (common in mathematics and statistics) that a
probability density vector multiplies ${\bf M}$ from the {\it left\/}. 
Therefore, in order for ${\bf M}$ to be a Markov matrix, each of its 
{\it row\/} sums must equal one. 
Given that the system is initially described by a
probability density represented by the row vector ${\vec v}_{\rm i}^{\rm T}$
(here taken to correspond to probability one for 
the metastable state with no overturned spins, transient state 0 in 
Fig.~\ref{fig:full}), the mean first-passage time to
the absorbing subspace can be shown to be\cite{IOSI80,NOVO01} 
\begin{equation}
\label{EAMCtau}
\langle\tau\rangle  
= 
{\vec v}_{\rm i}^{\rm T} \left( {\bf I} -{\bf T} \right)^{-1} {\vec e}
\; ,
\end{equation}
where $\vec e$ is a column vector with all elements equal to one, and
$\bf I$ is the $t\times t$ unit matrix.
The transient and absorbing states are shown in Fig.~\ref{fig:full}. 
The transition rates that form the elements of $\bf T$ were calculated
using the $p_i$ from Table~\ref{tab:prob} and degeneracies 
$n_i^g$ and $n_i^s$ deduced
from the geometric configurations shown in Fig.~\ref{fig:full}. 
Equation~(\ref{EAMCtau}) was then solved analytically for 
$\beta \rightarrow \infty$ 
in a computer-aided calculation using Mathematica.\cite{WOLF96} 
The metastable lifetime was calculated in this way, 
both at the special values of $|H|$ where
the one-step analytical calculations indicated that $A$ should have
discontinuities ($|H| = 1.5$, 2.0, 2.5, 3.5, and 4.5 for the TDA
dynamic, and $|H| = 2.0$ and 3.0 for the OSD dynamic), and at several
points in the field intervals where $A$ was expected to be constant. 
The results of these analytic calculations are included in 
Tables~\ref{tab:TDAAG} and~\ref{tab:BVAG} and also shown in
Figs.~\ref{fig:TDA} and~\ref{fig:OSDgam}.
The only field value where differences were found between the one-step
and full analytic calculations was $|H|=2$ for both dynamics. 

\subsection{Monte Carlo simulations}
\label{sec:MC}
Both sets of analytic results were further checked by 
continuous-time Monte Carlo simulations with time measured in MCS as
elsewhere in this paper. For $|H| < 4.5$ we used 
the Monte Carlo with absorbing Markov chains 
(MCAMC) method,\cite{NOVO02,NOVO95,NOVO01,NOVO03B} 
which significantly speeds up the simulations while faithfully 
preserving the underlying dynamic. The MCAMC method was applied with 
two transient
states (states T0 and T1 in Fig.~\ref{fig:full}). After exit from 
this transient
subspace, the simulation continued with the standard $n$-fold way 
algorithm\cite{BORT75,NOVO95} until the magnetization reached zero. 
At the lowest
temperatures the MPFUN arbitrary-precision package\cite{BAIL90,BAIL95} 
was used to 
obtain sufficient numerical precision in the inversion of 
$({\bf I} - {\bf T})$ 
in Eq.~(\ref{EAMCtau}), which in this case is a $2 \times 2$ matrix. 
For $|H| \ge 4.5$, where the lifetime is the average first-passage
time to a single overturned spin, we used the $n$-fold way algorithm,
needing only to consider the transitions out of class~0 with rate $Np_1$.
The system was an $L=24$ square with periodic boundary conditions, 
and 2000 escapes were used (100\,000 for $|H| \ge 4.5$).
Simulations were performed, both at the same special values of $|H|$ that 
were studied by the computer-aided analytic AMC method, and at several
points in the field intervals where $A$ was expected to be constant. 
The system was started in the metastable state with all spins up and
subjected to a negative field of magnitude $|H|$, and the
metastable lifetime was defined as the average time until the 
system magnetization reached zero (except for $|H| \ge 4.5$, where it was
defined as the average time until a single spin was overturned). The
lifetime was then measured at each $|H|$ for a series of temperatures
between $T = 0.08$ and a field-dependent minimum value (between
approximately 0.04 for the smallest $|H|$ and about 0.003 for the
largest $|H|$). Some of the results, shown vs $T$ as $T \ln \langle
\tau \rangle \approx \Gamma + T \ln A$, 
are plotted in Fig.~\ref{fig:all} for the TDA dynamic. 
To find $\Gamma$ and $A$, linear fits to such plots were performed, deleting
higher-$T$ points (which may deviate from the low-temperature 
linear $T$-dependence) 
until the $Q$-factor of the fit became acceptable.\cite{NUMRECF} 
The results of these simulations are included in 
Tables~\ref{tab:TDAAG} and~\ref{tab:BVAG} and also shown in
Figs.~\ref{fig:TDA} and~\ref{fig:OSDgam}.
The agreement with the analytical results is everywhere excellent. 
At $|H|=2$, where there are differences between the one-step and
full analytical results for $A$, the simulated results  
agree with the full analytical ones to within one standard deviation for both 
dynamics.

\section{Summary and Conclusions}
\label{sec:conc}

In this work we have continued our studies of the influence of different
kinetic Monte Carlo transition rates on the simulated
structure and mobility of
interfaces\cite{RIKV00B,RIKV02,RIKV02B,RIKV03} 
and on the low-temperature simulated nucleation 
properties\cite{PARK02B,PARK02C,PARK02,PARK04} of Ising and lattice-gas
models. These studies have documented that 
different transition rates can lead to very significant differences in
both structure and dynamics, even between dynamics that all satisfy detailed
balance and also have the same conservation properties. 
As an extension of these studies, we here considered two popular
dynamics that include a local energy barrier representing a transition
state inserted between individual Ising or
lattice-gas states. Such Arrhenius dynamics, as they are often called,
are appropriate in kinetic Monte Carlo simulations of discrete Ising or
lattice-gas models in which the discrete states serve as approximations
for high-probability configurations in an underlying continuous
potential. Examples are the use of a lattice-gas model to study
diffusion in a continuous corrugation 
potential,\cite{ALAN92A,ALAN92,ALAN02,MITC02} or the use
of an Ising model as an approximation for a continuum spin model with a
strong uniaxial anisotropy.\cite{MUNO03} 

The two Arrhenius dynamics that we studied are the commonly used
one-step dynamic (OSD)\cite{KANG89,FICH91,FICH02} [Eq.~(\ref{eq:OSD})]
and the two-step
transition dynamics approximation (TDA)\cite{ALAN92A,ALAN92,ALAN02} 
[Eq.~(\ref{eq:TDA})]. 
Using the same analytical and numerical methods that we used 
in Ref.~\onlinecite{PARK04} for the standard Glauber dynamic [Eq.~(\ref{eq:G})] 
and the soft Glauber dynamic [Eq.~(\ref{eq:SG})], we calculated the
exponential argument $\Gamma$ and prefactor $A$ in Eq.~(\ref{eq:tau})
for the metastable lifetime $\langle \tau \rangle$ (the inverse of the
nucleation rate per unit volume) in nucleation simulations at very low
temperatures. (Although the dynamics studied are different, the more
detailed descriptions of the methods given here can also serve as
supplementary material for the necessarily brief descriptions in
Ref.~\onlinecite{PARK04}.) 
While the differences between the OSD and TDA dynamics,
and between them and the standard Glauber dynamic, are not as dramatic
as those between the standard and soft Glauber
dynamics in Ref.~\onlinecite{PARK04}, they are definitely significant. Once
again they confirm that, contrary to what has been a common part of the
folklore of nucleation theory, $\Gamma$ does {\it not\/} always equal
the height of the energy barrier separating the metastable state from
the stable state. The differences that we find between the results for
the OSD and TDA rates also indicate that these dynamics are not
physically equivalent. It is thus interesting to note that
differences have also been observed between the TDA and a version of the
OSD in which $E_{\rm T}$ is constant (called the initial-value
approximation, IVA) in
simulations of diffusion in a lattice-gas model of oxygen on the (100)
surface of tungsten at elevated 
temperatures.\cite{VATT97,UEBI98C,VATT98C,UEBI98,VATT99B,UEBI99} 

One way of obtaining the correct dynamics for a given system could
be to start from the quantum-mechanical formulation, and from that
derive the dynamic for the classical model within the appropriate
approximations relevant for the particular system.  For example, one can
utilize the time-dependent quantum density matrix for a Hamiltonian
coupled to a heat bath, integrate out the heat bath, and with additional
approximations usually arrive at a dynamic for a classical or
semiclassical system.  This was accomplished for the quantum spin-1/2
system with a fermion heat-bath, giving the Ising model with the standard, hard
Glauber dynamic.\cite{MART77}  It has recently also been accomplished
for a quantum spin-1/2 lattice system coupled to a bosonic heat bath,
giving a dynamic that is different from the hard Glauber 
dynamic.\cite{PARK02C,PARK02B,PARK02} 

In this paper we have only studied the effect that changing the dynamic has 
on the two-dimensional square-lattice Ising model.  
It is natural to ask how widely such
influences on nucleation occur in other models and in other dimensions.  At
low temperatures, studies have also been made of the three-dimensional Ising
model, both through simulations \cite{NOVO03} and through rigorous
mathematical analysis.\cite{BOVI02,C1,C2,BA1}  (Note that the results
given for three-dimensional systems in
Refs.~\onlinecite{BOVI02,C1,C2} are not correct for all $|H|$; 
correct results are given in Refs.~\onlinecite{BA1,NOVO03}.)  
We expect that both $\Gamma$ and the prefactor
should also change for different dynamics for the cubic-lattice Ising
model.  In fact, we anticipate that results similar to those presented
here may hold for most lattice models with discrete state spaces.  One
important question is whether in off-lattice simulations, for
example ones that use energy barriers from zero-temperature first-principles
calculations,\cite{MITC02} 
a change in the dynamic could also change both $\Gamma$
and the prefactor in low-temperature nucleation.  If this is the case,
then, since a small change in $\Gamma$ can lead to orders-of-magnitude
changes in lifetimes, particular attention must be paid to deriving
the dynamic to be used in off-lattice simulations.

The results of this study reinforce the message that great caution must be
exercised when constructing transition rates for kinetic Monte Carlo
simulations of specific physical
systems.\cite{RIKV00B,RIKV02,RIKV02B,RIKV03,PARK04}
On the other hand, the
significant differences between individual dynamics that have been
observed indicate that experimental results for such quantities as
nucleation rates and interface mobilities should provide valuable input
for devising physically sound kinetic Monte Carlo algorithms in the
future.

\begin{acknowledgments}
G.M.B.\ gratefully acknowledges the hospitality of the Mississippi 
State University Department of Physics and Astronomy and ERC Center for 
Computational Sciences, and of the Florida State University School of 
Computational Science and Information Technology and Center for Materials 
Research and Technology. 
This work was supported by NSF grants No.\ DMR-0120310 and DMR-0240078. 
\end{acknowledgments}


\begin{thebibliography}{10}
\bibitem{NOTE1}
  Starting from the Ising Hamiltonian of Eq.~(\ref{eq:ham}), the standard,
  explicit mapping between the Ising and lattice-gas formulations is as
  follows. We identify the Ising variable $\sigma_\alpha = +1$ (``spin up'') at
  site $\alpha$ with the lattice-gas variable $c_\alpha = 1$ (occupied or
  ``solid'') and $\sigma_\alpha =-1$ with $c_\alpha =0$ (empty or ``fluid''),
  so that $\sigma_\alpha = 2 c_\alpha -1$. The Ising and lattice-gas
  interaction constants [$J$ and $\phi$ respectively; $J$ is set equal 
  to unity in Eq.~(\ref{eq:ham})] are related as $\phi = 4
  J$, and the applied Ising field $H$ is related to the lattice-gas chemical
  potential $\mu$ as $H = (\mu - \mu_0)/2$, where $\mu_0 = -8J = - 2 \phi$ is
  the coexistence value of $\mu$.
\end{thebibliography}

\clearpage

\begin{table}
\caption[]{
Rates of flipping a spin in class $m$, $p_m$, for $|H|>0$ and $\Delta \ge
0$ in the limit 
of $\beta \rightarrow \infty$. Here $\sigma = +$ ($-$) corresponds to a spin 
in the metastable (stable) direction, and $N_+$ is
the number of its nearest-neighbor spins in the metastable direction.
The analytical forms of the OSD results do not depend on $|H|$. 
The TDA results marked $\ast$ all take the value $1 \over 4$ for $\Delta = 0$. 
The first two lines of TDA results for classes 4 and 9 are only applicable for $\Delta > 2$, and 
the first two lines of TDA results for classes 5 and 10 are only applicable for $\Delta > 4$. 
} 
\label{tab:prob}
\begin{tabular}{c|c|c|c|c||c|c|c|c|c}
$m$ & $\sigma$ & $N_+$ & $p_m^{\rm OSD}$   & $p_m^{\rm TDA}$
& $m$ & $\sigma$ & $N_+$ & $p_m^{\rm OSD}$ & $p_m^{\rm TDA}$  \\ \hline
1   & +   & 4     & $e^{-\beta (\Delta+4-|H|)}$ & $e^{-\beta (8-2|H|)}$ for $0 < |H|< 4 - \Delta$
& 6 & $-$ & 4     & $e^{-\beta (\Delta-4+|H|)}$     & 1 for $0 < |H|< 4-\Delta$  \\ 
    &     &       &                       & $\frac{1}{2}e^{-2 \beta \Delta}$ for $|H|=4-\Delta$ $\ast$ 
&   &     &       &                       & $1 \over 2$ for $|H|=4-\Delta$ $\ast$ \\  
    &     &       &                       & $e^{-\beta (\Delta+4-|H|)}$ for $4-\Delta<|H|<4$
&   &     &       &                       & $e^{-\beta (\Delta-4+|H|)}$ for $4-\Delta<|H|<4$\\  
    &     &       &                       & $e^{-\beta \Delta}$ for $|H|=4$ $\ast$
&   &     &       &                       & $e^{-\beta \Delta}$ for $|H|=4$ $\ast$\\  
    &     &       &                       & $e^{-\beta (\Delta+4-|H|)}$ for $4<|H|<4+\Delta$
&   &     &       &                       & $e^{-\beta (\Delta-4+|H|)}$ for $4<|H|<4+\Delta$\\  
    &     &       &                       & $1 \over 2$ for $|H|=4+\Delta$ $\ast$
&   &     &       &                       & $\frac{1}{2}e^{-2 \beta \Delta}$ for $|H|=4+\Delta$ $\ast$ \\  
    &     &       &                       & 1 for $|H|>4+\Delta$
&   &     &       &                       & $e^{-\beta (-8+2|H|)}$ for $|H|>4+\Delta$ 
                                                                                   \\ \hline 
2   & +   & 3     & $e^{-\beta (\Delta+2 -|H|)}$ & $e^{-\beta (4-2|H|)}$ for $0 < |H|< 2-\Delta$
& 7 & $-$ & 3     & $e^{-\beta (\Delta-2 +|H|)}$ & 1 for $0 < |H|< 2-\Delta$ \\
    &     &       &                              & $\frac{1}{2}e^{-2 \beta \Delta}$ for $|H|= 2-\Delta$ $\ast$
&   &     &       &                              & $1 \over 2$ for $|H|= 2-\Delta$ $\ast$ \\
    &     &       &                              & $e^{-\beta (\Delta+2 -|H|)}$ for $2-\Delta < |H| < 2$
&   &     &       &                              & $e^{-\beta (\Delta-2 +|H|)}$ for $2-\Delta < |H| < 2$\\
    &     &       &                              & $e^{-\beta \Delta}$ for $|H|=2$ $\ast$ 
&   &     &       &                              & $e^{-\beta \Delta}$ for $|H|=2$ $\ast$ \\
    &     &       &                              & $e^{-\beta (\Delta +2 -|H|)}$ for $2<|H|<2+\Delta$
&   &     &       &                              & $e^{-\beta (\Delta -2 +|H|)}$ for $2<|H|<2+\Delta$ \\
    &     &       &                              & $1 \over 2$ for $|H|=2+\Delta$ $\ast$
&   &     &       &                              & $\frac{1}{2}e^{-2 \beta \Delta}$ for $|H|=2+\Delta$ $\ast$ \\
    &     &       &                              & 1 for $|H|>2+\Delta$
&   &     &       &                              & $e^{-\beta (-4+2|H|)}$ for $|H|>2+\Delta$ 
                                                                      \\ \hline
3   & +   & 2     & $e^{-\beta (\Delta -|H|)}$   & $e^{-\beta (\Delta -|H|)}$ for $0<|H|<\Delta$
& 8 & $-$ & 2     & $e^{-\beta (\Delta +|H|)}$   & $e^{-\beta (\Delta +|H|)}$ for $0<|H|<\Delta$\\ 
    &     &       &                              & $1 \over 2$ for $|H|=\Delta$ $\ast$
&   &     &       &                              & $\frac{1}{2}e^{- 2 \beta \Delta }$ for $|H|=\Delta$ $\ast$ \\ 
    &     &       &                              & 1 for $|H|>\Delta$
&   &     &       &                              & $e^{- 2 \beta |H|}$ for $|H|>\Delta$
                                                                      \\ \hline
4   & +   & 1     & $e^{-\beta (\Delta-2-|H|)}$  &  $e^{-\beta (\Delta-2-|H|)}$ for $0<|H|<\Delta-2$ 
& 9 & $-$ & 1     & $e^{-\beta (\Delta+2+|H|)}$  &  $e^{-\beta (\Delta+2+|H|)}$ for $0<|H|<\Delta-2$ \\
    &     &       &                              &  $1 \over 2$ for $|H|=\Delta-2$ 
&   &     &       &                              &  $\frac{1}{2}e^{-2 \beta \Delta}$ for $|H|=\Delta-2$ \\
    &     &       &                              &  1 for $|H| > \Delta-2$ 
&   &     &       &                              &  $e^{-\beta (4+2|H|)}$ for $|H| > \Delta-2$
                                                                      \\ \hline
5   & +   & 0     & $e^{-\beta (\Delta-4-|H|)}$  &  $e^{-\beta (\Delta-4-|H|)}$ for $0 < |H| < \Delta-4$
&10 & $-$ & 0     & $e^{-\beta (\Delta+4+|H|)}$  &  $e^{-\beta (\Delta+4+|H|)}$ for $0 < |H| < \Delta-4$ \\
    &     &       &                              &  $1 \over 2$ for $|H| = \Delta-4$
&   &     &       &                              &  ${1 \over 2} e^{-2 \beta \Delta}$ for $|H| = \Delta-4$ \\
    &     &       &                              &  1 for $|H| > \Delta-4$
&   &     &       &                              &  $e^{-\beta (8+2|H|)}$ for $|H| > \Delta-4$
                                                                      \\ 
\end{tabular}
\end{table}

\clearpage

\begin{table}
\caption[]{
Analytical and simulated $A_{\rm TDA}$ and $\Gamma_{\rm TDA}$ 
for the TDA dynamic with $|H| > 1$.
Theoretical results are valid for $0 \le \Delta < 1$, 
while all numerical values are for $\Delta = 0.5$. 
For $A_{\rm TDA}$, the analytical AMC calculation with a one-step Markov
chain and the full AMC calculation may give different results for
fields $H$ such that $2/|H|$ is a positive integer, 
as well as for $|H| \pm \Delta$.  They are therefore given as 
$A_{\rm TDA}$(1-step) and $A_{\rm TDA}$(full), respectively. 
For $\Gamma_{\rm TDA}$, the two analytical methods always
agree, and the results are given together as $\Gamma_{\rm TDA}$(anal.). 
The simulated results, $A_{\rm TDA}$(sim.) and $\Gamma_{\rm TDA}$(sim.),
are obtained from weighted two-parameter fits to 
plots of $T \ln \langle \tau \rangle \approx \Gamma + T \ln A$, such
as those shown 
in Fig.~\protect\ref{fig:all}. Standard errors in the last
digit are given in parentheses. 
}
\label{tab:TDAAG}
\begin{tabular}{c|c|c|c|c|c}
$|H|$ & $A_{\rm TDA}$(1-step) & $A_{\rm TDA}$(full) & $A_{\rm TDA}$(sim.) 
& $\Gamma_{\rm TDA}$(anal.) & $\Gamma_{\rm TDA}$(sim.) 
                                                                      \\ \hline
$1<|H|<2-\Delta$ & $3 \over 8$ = 0.375 & $3 \over 8$ = 0.375 
                        &  & $16-6|H|$ &  
                                                                      \\ \hline
$|H|=2-\Delta = 1.5$   & $1 \over 2$ = 0.5 & $1 \over 2$ = 0.5 
                        &  & $6 \Delta+4 = 7$ &  
                                                                      \\ \hline
$2-\Delta<|H|<2$ & $1 \over 8$ = 0.125 & $1 \over 8$ = 0.125 
                        & 0.11(1) at $|H|=1.8$ & $\Delta +14-5|H| = 14.5-5|H|$ & 5.506(5) at $|H|=1.8$ \\
                 &                     &  
                        & 0.11(1) at $|H|=1.95$ &                              & 4.753(2) at $|H|=1.95$ 
                                                                      \\ \hline
$|H|=2$ & $3 \over 8$ = 0.375 & ${330 \over 997} \approx 0.33099$ & 0.335(6)  & $\Delta+4=4.5$ & 4.4998(7) 
                                                                      \\ \hline
$2<|H|<2+\Delta$ & $1 \over 4$ = 0.25 & $1 \over 4$ = 0.25 &  0.246(4) at $|H|=2.25$ 
                & $\Delta+10-3|H| = 10.5-3|H|$ & 3.7498(3) at $|H|=2.25$ \\ \hline
$|H|=2+\Delta=2.5$ & $3 \over 2$ = 1.5 & $3 \over 2$ = 1.5 & 1.45(3)
                & $4-2\Delta=3$ &  3.0000(7)           \\ \hline
$2+\Delta<|H|<4-\Delta$ & $5 \over 4$ = 1.25 & $5 \over 4$ = 1.25 & 1.21(2) at $|H|=3$
                & $8-2|H|$ &  2.0000(7) at $|H|=3$   \\ \hline
$|H|=4-\Delta=3.5$ & $9 \over 4$ = 2.25 & $9 \over 4$ = 2.25 & 2.19(4)
                & $2\Delta=1$ & 0.9999(7)  \\ \hline
$4-\Delta<|H|<4$ & 1 & 1 & 0.98(2) at $|H|=3.75$ & $\Delta+4-|H|=4.5-|H|$ & 0.7498(3) at $|H|=3.75$\\ \hline
$|H| = 4$ & 1   & 1 & 0.97(2)  & $\Delta=0.5$ & 0.5000(1)  \\ \hline
$4<|H|<4+\Delta$ & 1 & 1 & 0.97(2) at $|H|=4.25$ & $\Delta+4-|H|=4.5-|H|$ &0.24998(8) at $|H|=4.25$\\ \hline
$|H|=4+\Delta = 4.5$   & 2   &   & 1.996(2) & 0 & 0.00006(3) \\ \hline
$|H|>4+\Delta$   & 1   &   & 1.001(1) at $|H|=4.75$  & 0 &  0.00001(3) at $|H|=4.75$   \\ \hline
\end{tabular} \end{table}

\begin{table}
\caption[]{
Analytical and simulated $A_{\rm OSD}$ and $\Gamma_{\rm OSD}$ for the 
OSD dynamic with $|H| > 1$. 
Theoretical results are valid for $0 \le \Delta < 1$, 
while all numerical values are for $\Delta = 0.5$. 
The estimation procedures and the meanings of all symbols are the same
as in Table~\protect\ref{tab:TDAAG}. 
}
\label{tab:BVAG}
\begin{tabular}{c|c|c|c|c|c}
$|H|$ & $A_{\rm OSD}$(1-step) & $A_{\rm OSD}$(full) & $A_{\rm OSD}$(sim.) 
                        & $\Gamma_{\rm OSD}$(anal.) & $\Gamma_{\rm OSD}$(sim.) 
                                                                      \\ \hline
$1<|H|<2$ & ${1 \over 8}$ = 0.125 & ${1 \over 8}$ = 0.125 & 0.121(7) at $|H|=1.8$ 
                        & $\Delta + 14 - 5|H| =14.5-5|H|$ & 5.500(3) at $|H|=1.8$ \\ 
          &             &                                 & 0.120(5) at $|H|=1.9$ 
                        &                                 & 5.000(2) at $|H|=1.9$ 
                                                                      \\ \hline
$|H|=2$   & ${3 \over 8} = 0.375$ & ${55 \over 162} \approx 0.3395 $ & 0.333(6) 
                        &$\Delta+4=4.5$& 4.5000(7) 
                                                                      \\ \hline
$2<|H|<3$ & ${1 \over 4}=0.25$ & ${1 \over 4}=0.25$ & 0.242(5) at $|H|=2.5$
                        & $\Delta+10-3|H| = 10.5-3|H|$ & 3.0000(7) at $|H|=2.5$ \\ \hline
$|H|=3$ & ${5 \over 4}=1.25$ & ${5 \over 4}=1.25$ & 1.21(2)
                        & $\Delta+1 = 1.5$ & 1.5000(7) \\ \hline
$|H|>3$ & 1 & 1 & 0.97(2) at $|H|=3.5$ & $\Delta+4-|H|=4.5-|H|$  & 1.0000(7) at $|H|=3.5$ \\ 
        &   &   & 1.000(1) at $|H|=4.75$ &                       & $-0.25002(3)$ at $|H|=4.75$ \\ \hline
\end{tabular}
\end{table}

\clearpage

\clearpage 

\begin{figure}[ht]
\includegraphics[angle=0,width=.47\textwidth]{ECgamma.eps}
\hspace{0.5truecm}
\includegraphics[angle=0,width=.47\textwidth]{ECpre.eps}
\vspace{.3cm}
\caption[]{
Analytical and simulated results for $\Gamma$ (a) and $A$ (b) vs $|H|$ for
the OSD dynamic, using $\Delta = 0.5$. Results for $\Gamma$ for
the standard Glauber dynamic (from Ref.~\protect\onlinecite{PARK04}) 
are shown by dashed lines in (a). 
See discussion in the text. 
}
\label{fig:OSDgam}
\end{figure}

\end{document}